\documentclass{article}
\usepackage{ijcai25}

\usepackage{times}
\usepackage{soul}
\usepackage{url}
\usepackage[hidelinks]{hyperref}
\usepackage[utf8]{inputenc}
\usepackage[small]{caption}
\usepackage{graphicx}
\usepackage{amsmath}
\usepackage{amsthm}
\usepackage{booktabs}
\usepackage{algorithm}
\usepackage{algorithmic}
\usepackage[switch]{lineno}

\usepackage{amsmath}
\usepackage{amssymb}
\usepackage{booktabs}
\usepackage{subcaption}
\usepackage[capitalise,noabbrev]{cleveref}
\usepackage{esvect}

\title{Quantifying the Self-Interest Level of Markov Social Dilemmas}

\author{
Richard Willis$^1$
\and
Yali Du$^1$\and
Joel Z Leibo$^{1,2}$\And
Michael Luck$^3$\\
\affiliations
$^1$King's College London\\
$^2$Google DeepMind\\
$^3$University of Sussex\\
\emails
richard.willis@kcl.ac.uk,
yali.du@kcl.ac.uk,
jzl@deepmind.com,
michael.luck@sussex.ac.uk
}

\begin{document}

\maketitle

\begin{abstract}
This paper introduces a novel method for estimating the self-interest level of Markov social dilemmas.
We extend the concept of self-interest level from normal-form games to Markov games, providing a quantitative measure of the minimum reward exchange required to align individual and collective interests.
We demonstrate our method on three environments from the Melting Pot suite, representing either common-pool resources or public goods.
Our results illustrate how reward exchange can enable agents to transition from selfish to collective equilibria in a Markov social dilemma.
This work contributes to multi-agent reinforcement learning by providing a practical tool for analysing complex, multistep social dilemmas.
Our findings offer insights into how reward structures can promote or hinder cooperation, with potential applications in areas such as mechanism design.
\end{abstract}

\section{Introduction}
\label{sec:introduction}

Social dilemmas are situations where individual incentives conflict with group interests, presenting significant challenges in multi-agent cooperation.
The main difficulty of social dilemmas is that prosocial actions can be personally costly.
Agents need sufficient motivation to care about others for collective action to become more attractive than selfish behaviour.
We address this with reward exchange, whereby agents agree to exchange a fixed proportion of their rewards with each other, creating an incentive for them to improve the well-being of others.

The self-interest level \cite{willis24__resolving_social_dilemmas_with_minimal_reward_transfer} quantifies the greatest proportion of their own rewards that agents can retain while using reward exchange to resolve a social dilemma.
It serves as a solution to social dilemmas and a metric for players' propensity to cooperate, assessing the gap between individual and collective incentives.
A high level indicates that players can achieve socially optimal outcomes with minimal consideration for others' interests, while a low level indicates strong incentives for players to avoid prosocial behaviour.

While the self-interest level can be computed analytically in normal-form games, a new method is needed for more complex game structures. 
This paper presents a novel method for empirically estimating the self-interest level of Markov game representations of social dilemmas using multi-agent reinforcement learning (MARL).
Our approach begins by training independent agents to maximise their individual rewards, which commonly results in suboptimal collective outcomes in social dilemmas.
We then introduce reward exchange and progressively increase the amount shared between agents, identifying the critical threshold where these initially selfish policies transition to collective behaviours.

Our primary contributions are twofold: we present a novel quantitative method for determining the self-interest level, superseding the qualitative approach employed in previous work \cite{willis23__resolving_social_dilemmas_through_reward_transfer_commitments}; and we provide more comprehensive experimental results on three environments featuring larger numbers of agents from the Melting Pot suite \cite{leibo21__scalable_evaluation_of_multiagent_reinforcement_learning_with_melting_pot}.
This research contributes to the growing fields of Cooperative AI \cite{dafoe20__open_problems_in_cooperative_ai} and MARL in mixed-motive scenarios \cite{du23__a_review_of_cooperation_in_multiagent_learning}.

\section{Related Work}
\label{sec:related_work}
Classically, social dilemmas have been modelled as matrix games \cite{axelrod80__effective_choice_in_the_prisoners_dilemma,schelling73__hockey_helmets_concealed_weapons_and_daylight_saving}.
More recently, researchers have modelled social dilemmas using Markov games \cite{leibo17__multiagent_reinforcement_learning_in_sequential_social_dilemmas,hughes18__inequity_aversion_improves_cooperation_in_intertemporal_social_dilemmas}, which feature greater real-world complexities, including graded cooperativeness, agents with partial information about the state of the world, and decisions with temporally extended consequences.

Our work focuses on two prominent approaches.
The first approach develops game-theoretic metrics to quantify the amount of shared interest required to achieve socially optimal equilibria in mixed-motive games \cite{apt14__selfishness_level_of_strategic_games,elias10__sociallyaware_network_design_games,chen08__altruism_selfishness_and_spite_in_traffic_routing,chen11__the_robust_price_of_anarchy_of_altruistic_games,caragiannis10__the_impact_of_altruism_on_the_efficiency_of_atomic_congestion_games,willis24__resolving_social_dilemmas_with_minimal_reward_transfer}.
These contributions are limited to analytically tractable games.
Our contribution addresses this gap by extending these metrics to complex, computationally intractable games.

The second direction focuses on MARL methods for independent agents operating without centralised control or intrinsic motivations \cite{du23__a_review_of_cooperation_in_multiagent_learning}. 
Within this framework, reward transfers have emerged as a mechanism for promoting collective behaviours.
Baker \cite{baker20__emergent_reciprocity_and_team_formation_from_randomized_uncertain_social_preferences} employs reward transfer to represent inter-agent social preferences and study their impact on team formation.
Schmid et al. \cite{schmid23__learning_to_participate_through_trading_of_reward_shares} introduce a marketplace mechanism to trade reward shares, allowing agents to acquire stakes in each other.
Gemp et al. \cite{gemp22__d3c} use reward transfers to minimise the social costs of local Nash equilibria.

Although these approaches provide valuable qualitative insights, they lack rigorous methods for determining the minimum reward transfers required to achieve socially optimal outcomes.
Our work addresses this limitation by establishing a quantitative threshold for the required amount of reward exchange to resolve a social dilemma, thereby adding a descriptive dimension to the analysis and minimising transfer costs when such transfers are costly.

Several studies have explored dynamic reward transfer mechanisms via additional game actions.
Lupu and Precup \cite{lupu20__gifting_in_multiagent_reinforcement_learning} and Wang et al. \cite{wang21__emergent_prosociality_in_multiagent_games_through_gifting} investigate gift-giving actions in tragedy of the commons and coordination games, respectively.
Yang et al. \cite{yang20__learning_to_incentivize_other_learning_agents} optimise reward transfers to shape the behaviour of learning algorithm opponents, thereby enhancing the overall reward for the transferee.
Yi et al. \cite{yi22__learning_to_share_in_multiagent_reinforcement_learning} enable agents to support collective behaviours by dynamically exchanging rewards with others nearby. 
With contracts, payments can be made conditional on specific joint actions being taken \cite{christoffersen22__get_it_in_writing}.
A fair value for an action has been proposed \cite{sodomka13__cocoq}.

However, action-level reward transfers require strategies to specify transfers for all joint actions across all states, increasing computational and strategic complexity.
Our approach simplifies this by requiring a single advance commitment to a fixed proportion to exchange.

\section{Background}
\label{sec:background}

\subsection{Reinforcement Learning in Markov Games}

A Markov game is played by $n$ players within a finite set of states $\mathcal{S}$.
The game is parameterised by sets of available actions for each player $\vv{\mathcal{A}} = (\mathcal{A}_1, \ldots, \mathcal{A}_n)$, and a stochastic transition function \(T : \mathcal{S} \times \vv{\mathcal{A}} \rightarrow \Delta (\mathcal{S})\), mapping from joint actions at each state to the set of discrete probability distributions over states.
A Markov game is partially observable if, instead of the state, the agents only view observations, $\vv{o}_t$, provided by the observation function $O(s_t)$.
After a transition, each agent $i$ receives a reward specified by their per-timestep reward function, $r_i(\vv{a}_t, s_t)$, where $\vv{a}_t \in \vv{\mathcal{A}}$ is the joint action at timestep $t$ and $s_t \in \mathcal{S}$ is the state at timestep $t$.
Each agent $i$ independently learns a policy $\pi_i(a_i | o_i)$ to maximise a long-term \emph{\(\gamma\)}-discounted episode reward, defined as:
\begin{equation}
\nonumber
R_i(\vv{\pi}) = \mathbb{E}_{\vv{a}_t \sim \vv{\pi}} \left[ \sum_{t=0}^{\infty} \gamma^t r_i(\vv{a}_t, s_t) \right]
\end{equation}

\subsection{Markov Social Dilemmas}
\label{sec:markov_sd}
Social dilemmas are situations in which individuals must choose between acting selfishly (to defect) for personal gain or acting in a prosocial manner (to cooperate), yielding greater overall benefits for the collective.
For all agents: (i) the collective does better when an agent chooses to cooperate than when the agent chooses to defect; (ii) each agent may be better off individually when it defects; and (iii) all agents prefer mutual cooperation over mutual defection.
We assess the impact of actions on the group using a social welfare metric, which quantifies a notion of collective good.
Let $\vv{R}(\vv{\pi}) = (R_1(\vv{\pi}), \ldots, R_n(\vv{\pi}))$ denote the tuple of episode rewards for all agents under joint policy $\vv{\pi}$.
As is common in the literature \cite{anshelevich04__the_price_of_stability_for_network_design_with_fair_cost_allocation,koutsoupias09__worstcase_equilibria,elkind20__price_of_pareto_optimality_in_hedonic_games}, we use utilitarian welfare $U$, which measures the unweighted sum of rewards obtained by all players.

\begin{equation}
  U(\vv{R}) = \sum_{i=1}^n R_i
  \label{eqn:utilitarian}
\end{equation}
An \emph{n}-player Markov social dilemma is a tuple \((M, \vv{\Pi} = \vv{\Pi}_c \cup \vv{\Pi}_d)\), where $M$ is a Markov game, and $\vv{\Pi}_c$ and $\vv{\Pi}_d$ are two disjoint sets of policies said to implement cooperation and defection, respectively.
The episode rewards satisfy the following properties:
\begin{enumerate}
  \item The utilitarian welfare is greater if an agent cooperates
  \begin{equation}
    \label{eqn:sd_ie_1}
    \forall i \quad U(\vv{R}({\pi_c}^\frown \vv{\pi_{-i}})) > U(\vv{R}({\pi_d}^\frown \vv{\pi_{-i}}))
  \end{equation}
  \item There is at least one joint policy profile for each agent where they are better off choosing to defect
  \begin{equation}
    \nonumber
    \forall i\ \; \exists \vv{\pi_{-i}} : R_i({\pi_d} ^\frown \vv{\pi_{-i}}) > R_i({\pi_c} ^\frown \vv{\pi_{-i}})
  \end{equation}
  \item All agents prefer mutual cooperation to mutual defection
  \begin{equation}
    \nonumber
    \forall i \quad R_i((\pi_c,\pi_c \ldots \pi_c)) > R_i((\pi_d,\pi_d \ldots \pi_d))
  \end{equation}
\end{enumerate}

Where $\vv{\pi_{-i}}$ represents the tuple of policies for all players other than player $i$, and $^\frown$ is a coupling operator that inserts $\pi_i$ into $\vv{\pi_{-i}}$ such that $\vv{\pi} = {\pi_i} ^\frown \vv{\pi_{-i}}$.

This technique of restricting the action space to a choice between fixed policies is known as empirical game-theoretic analysis \cite{walsh02__analyzing_complex_strategic_interactions_in_multiagent_systems,wellman06__methods_for_empirical_gametheoretic_analysis}.
In practice, we use learning algorithms to discover the policies.

\subsection{Reward Exchange}
\label{sec:reward_transfer}

We allow agents to enter into a contract to exchange portions of their future rewards among each other.
Without reward transfers, each agent is self-interested, aiming to maximise their personal reward.
Sharing rewards incentivises the agents to consider the impact of their actions on the other agents.

We introduce a parameter, $s$, which governs the proportion of its own rewards that an agent retains, termed the \emph{self-interest} of the agents.
The remainder, $1-s$, is distributed equally among the other $n-1$ co-players.
The post-transfer reward for agent $i$ comprises the retained portion of its own episode reward, $R_i$, plus any reward received from others:
\begin{equation}
  \nonumber
  R_i'(\vv{R}, s) = s R_i + \frac{1-s}{n-1}\sum_{j \neq i} R_j
\end{equation}

\subsection{Self-Interest Level}
\label{sec:self_interest_level}

We say that a social dilemma is resolved when all agents prefer to cooperate.
The self-interest level of a Markov social dilemma, denoted $s^*$, represents the maximum amount of self-interest that agents can retain while using reward exchange to resolve the dilemma, and is defined as:
\begin{equation}
\nonumber
s^* = \max \{s \mid \forall i \; R_i'(\vv{R}({\pi_c} ^\frown \vv{\pi_{-i}}), s) > R_i'(\vv{R}({\pi_d} ^\frown \vv{\pi_{-i}}), s)\}
\end{equation}
Note that when $s=\frac{1}{n}$, the post-transfer reward function for all agents is equivalent to maximising the utilitarian metric (\cref{eqn:utilitarian}), also referred to as a collective or team reward:
\begin{equation}
  \nonumber
  R_i'(\vv{R},\frac{1}{n}) = \frac{1}{n} \sum_{j=1}^n{R_j} = \frac{1}{n} U(\vv{R})
\end{equation}
Consequently, in a Markov social dilemma with $s=\frac{1}{n}$, cooperative policies are preferable to defect policies because cooperation increases the total reward due to \cref{eqn:sd_ie_1}.
However, players may still strictly prefer cooperative policies for $s > \frac{1}{n}$.
The self-interest level, therefore, has a lower bound, and $s^* \in [\frac{1}{n},1]$.

\paragraph{Example}
Consider the Public Goods Game where $n$ players each have \$1 and can contribute any amount $c \in [0,1]$ to a public fund.
Individual contributions up to $c_x$ are multiplied by $k_1$, while those above are multiplied by $k_2$ (where $1 < k_2 < k_1 < n$).
The payoff to player $i$ contributing $c_i$ is $1 - c_i + \frac{1}{n}\sum_{j=1}^n (k_1 \min(c_x, c_j) + k_2 \max(0, c_j - c_x))$.
Without reward exchange, contributing nothing dominates.
With reward exchange, partial cooperation ($c = c_x$) emerges when $\frac{k_2}{n} < s < \frac{k_1}{n}$, while full cooperation ($c = 1$) requires $s < \frac{k_2}{n}$.

\section{Method}
\label{sec:method}

We present our method in two parts.
First, we describe our approach to estimating the self-interest level of Markov social dilemmas.
Our goal is a method that takes a Markov game and outputs its self-interest level.
Then, we provide practical details, including environments and training procedures.

\subsection{Estimating the Self-interest Level}

Cooperation is a graded quantity in Markov social dilemmas, making it challenging to assess the degree to which a policy cooperates.
We consider joint policies that achieve an equivalent collective reward to those trained to maximise the sum of rewards (when $s = \frac{1}{n}$) to be maximally cooperating.
The self-interest level is estimated to be the largest value of $s$ for which independent policies achieve at least the same social welfare as those with a team reward, implying that they are fully cooperating.

Writing joint policies trained with a self-interest of $s$ as $\vv{\pi_s}$:
\begin{equation}
\nonumber
s^* = \max_{\frac{1}{n} \leq s \leq 1} : U(\vv{R}(\vv{\pi_s})) \approx U(\vv{R}(\vv{\pi_\frac{1}{n}}))
\end{equation}

We wish to find strong evidence that policies trained at the self-interest level converge to cooperative equilibria, regardless of their initialisation, implying that cooperation is dominant.
This is achieved by choosing challenging initialisations: policies trained without reward exchange, so the agents have incentives to act selfishly and shirk cooperation.
We theorise that these policies are far from cooperative behaviours in the policy space.
If, after introducing reward exchange, the policies subsequently converge to cooperative equilibria, we can be reasonably confident that cooperative equilibria are the only attractors in the policy space.

\paragraph{Policy Training}
Mixed-motive environments are notoriously difficult for independent MARL \cite{du23__a_review_of_cooperation_in_multiagent_learning}, as agents treat their co-players as static, which can hinder coordination and the discovery of optimal cooperative joint policies.
Furthermore, MARL can struggle due to credit assignment problems, particularly when using a team reward, which may lead to so-called 'lazy' agents.
To mitigate these issues, we implement a curriculum learning approach \cite{bengio09__curriculum_learning}, which helps to capture the true strategic structure of the game rather than artefacts of poor policy choices:
\begin{enumerate}
  \item \textbf{Pretraining}:
  We gradually increase the number of players in the environment, starting with a single player and progressively adding players until reaching the maximum number.
  This allows policies to learn environmental dynamics before addressing increasingly complex multi-agent interactions.

  \item \textbf{Training}: 
  We continue training the independent policies while iteratively decreasing their self-interest.
  This gradual approach aims to alleviate potential catastrophic forgetting caused by the change in reward function.
\end{enumerate}

\paragraph{Evaluation}
Due to the stochastic nature of reinforcement learning and Markov games, we repeat the training for multiple policy initialisations.
Furthermore, we identify the self-interest level with a degree of tolerance, selecting the largest value of $s$ that achieves a total reward that is not statistically worse than the best measured:
\begin{enumerate}
  \item Compute the mean and standard deviation of the collective reward at the end of training for each value of $s$, and identify $s_{\text{max}}$ as the $s$ value with the largest mean.
  \item Conduct a one-sided Dunnett's test \cite{dunnett55__a_multiple_comparison_procedure_for_comparing_several_treatments_with_a_control}, a method to compare multiple samples with a single control, to assess which of the means are statistically worse than $s_{\text{max}}$.
  \item Choose $s^*$ as the largest $s$ value with a mean that is not statistically worse, otherwise $s^* = s_{\text{max}}$ if all are worse.
\end{enumerate}

\subsection{Environments}

We evaluate our approach using three environments from the Melting Pot suite \cite{leibo21__scalable_evaluation_of_multiagent_reinforcement_learning_with_melting_pot}: Commons Harvest, Clean Up, and Externality Mushrooms\footnote{Images: \href{https://github.com/willis-richard/meltingpot/blob/markov_sd/data/commons_harvest__open/commons_harvest__open.jpeg}{Commons Harvest}, \href{https://github.com/willis-richard/meltingpot/blob/markov_sd/data/clean_up/clean_up.jpeg}{Clean Up}, \href{https://github.com/willis-richard/meltingpot/blob/markov_sd/data/externality_mushrooms__dense/externality_mushrooms__dense.jpeg}{Externality Mushrooms}}.
Commons Harvest models a tragedy of the commons scenario with a finite common-pool resource, where the challenge is to avoid overexploitation.
Clean Up and Externality Mushrooms represent public goods problems, where agents can invest to improve resource quality at personal cost; the challenge here is to prevent under-provisioning.
Commons Harvest and Clean Up are standard benchmarks for evaluating algorithms in social dilemmas \cite{jaques19__social_influence_as_intrinsic_motivation_for_multiagent_deep_reinforcement_learning,schmid23__learning_to_participate_through_trading_of_reward_shares,mckee23__a_multiagent_reinforcement_learning_model_of_reputation_and_cooperation_in_human_groups,baker20__emergent_reciprocity_and_team_formation_from_randomized_uncertain_social_preferences,wang19__achieving_cooperation_through_deep_multiagent_reinforcement_learning_in_sequential_prisoners_dilemmas}, while Externality Mushrooms provides an additional test case with similar cooperative dynamics.
Full details of these environments can be found in \cite{leibo21__scalable_evaluation_of_multiagent_reinforcement_learning_with_melting_pot}.

\paragraph{Commons Harvest}
Commons harvest comprises seven agents that harvest apples from four large and two small apple patches.
Collecting an apple provides a reward of 1.
Harvested apples regrow with a probability proportional to the number of apples remaining in the patch.
However, if all the apples in a patch are harvested, it is depleted, and no apples will regrow.
As long as one apple remains, given sufficient time, all the apples will regrow.

\paragraph{Clean Up}
Seven agents harvest apples, each providing a reward of 1.
In this environment, apples grow only if the river is sufficiently clean.
Over time, the river accumulates pollution that the agents may clean.
Pollution reduces the rate of apple growth, and no apples will grow if the level of pollution reaches 40\% or greater.

\paragraph{Externality Mushrooms}
This environment has five agents and four types of mushrooms distributed throughout the map.
Red mushrooms yield a reward of 1 to the individual who eats them.
Green mushrooms provide a reward of 2, shared between all agents.
Blue mushrooms give a reward of 3, divided equally among all agents except the one who consumed them.
Orange mushrooms destroy red mushrooms.
Thus, red mushrooms represent self-interest, green mushrooms embody prosocial behaviour, blue mushrooms exemplify altruism, and orange mushrooms are a punishment mechanism. 
Notably, consuming mushrooms promotes the growth of new mushrooms of the same colour.

\subsection{Implementation Details}
\label{sec:implementation_details}

\paragraph{Learning Algorithm}
We choose Proximal Policy Optimisation (PPO) \cite{schulman17__proximal_policy_optimization_algorithms} for its ease of use and widespread adoption in the field.
We opt against using parameter sharing (also known as self-play) to enable different agents to implement different policies.
Our implementation utilises a shallow neural network containing an LSTM cell.

\paragraph{Environment Configuration}
For all environments, we fix the episode length to 2000 timesteps, and we modify the observation space by compressing each grid cell from 8x8 pixels to a single pixel.
In Clean Up, we scale the growth rates of apples and pollution by $\frac{n}{7}$ when played with $n$ players, to maintain the incentive structure.
This scaling is also necessary to enable the river to be cleaned: in the default setting, pollution increases faster than a single agent can clean.

\paragraph{Hyperparameter Tuning}
For each environment, we tune our hyperparameters using a single-player version, allowing us to identify the best parameters as those achieving the greatest reward, which is not necessarily the case in the multiplayer case due to the mixed-motive reward structure.
The exploration problem is particularly challenging in Clean Up, so we cap pollution at 40\% for hyperparameter tuning.
This ensures that any cleaning will cause some apples to spawn, rather than needing to clean for many consecutive timesteps.
We subsequently use the best policy as the initial policy for pretraining in Clean Up, ensuring that the policies have acquired cleaning and harvesting skills, whereas the other environments start with randomly initialised policies.

\paragraph{Training Specifications}
For our experiments, we use five random seeds and train for 9000 episodes (18 million environment steps) at each stage of the curriculum.
We use a range of self-interest values based on the ratio of the fraction of reward an agent keeps for itself compared to the proportion of a co-player's reward it receives, because the agents typically face a choice between taking a benefit for themselves or allowing a co-player to gain it.
The ratios we use are $[20\!\!:\!\!1, 10\!\!:\!\!1, 5\!\!:\!\!1, 3\!\!:\!\!1, 5\!\!:\!\!2, 2\!\!:\!\!1, 5\!\!:\!\!3, 4\!\!:\!\!3, 1\!\!:\!\!1]$.
We use a p-value threshold of 0.1 for the Dunnett's test.
See \url{https://github.com/willis-richard/meltingpot/tree/markov_sd} for further details.

\section{Results}
\label{sec:results}

\subsection{Pretraining}
\label{sec:pre_training}

\begin{figure}[t]
  \centering
  \begin{subfigure}[t]{\columnwidth}
    \centering
    \includegraphics[width=\columnwidth]{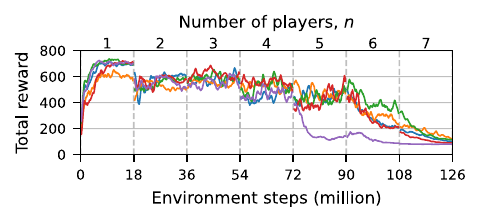}
    \vspace{-4ex}
    \caption{Commons Harvest}
    \label{fig:ch_pretraining}
  \end{subfigure}
  \begin{subfigure}[t]{\columnwidth}
    \centering
    \includegraphics[width=\columnwidth]{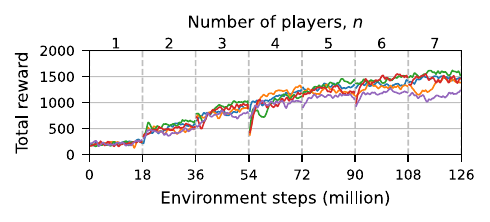}
    \vspace{-4ex}
    \caption{Clean Up}
    \label{fig:cu_pretraining}
  \end{subfigure}
  \begin{subfigure}[t]{\columnwidth}
    \centering
    \includegraphics[width=\columnwidth]{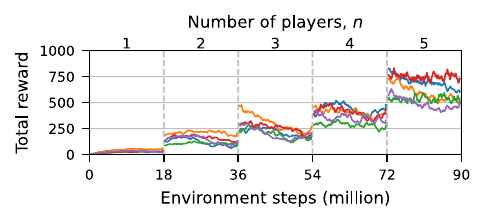}
    \vspace{-4ex}
    \caption{Externality Mushrooms}
    \label{fig:em_pretraining}
  \end{subfigure}
  \vspace{-1ex}
  \caption{Pretraining, increasing numbers of players}
  \vspace{-2ex}
\end{figure}

\paragraph{Commons Harvest}\cref{fig:ch_pretraining} shows that the best performance is achieved when there is only one agent.
In principle, multiple players should be able to match or exceed the reward of a single agent.
That they do not do so in practice is due to the difficulty of coordination and the mixed-motive structure of the rewards for $n > 1$ players.
The benefits of harvesting an apple are entirely captured by the harvester, but the future cost of a reduced regrowth rate is shared among all agents; the agents may therefore have incentives to harvest more apples than is socially optimal.

Over the range of 2--4 players, all five seeds maintain good social outcomes.
However, with 5--7 players, the performance collapses.
At this point, the personal cost of overharvesting has decreased sufficiently to tempt agents to over-consume.
This behaviour initially increases their reward, but as all policies follow suit, all agents are worse off.
In these equilibria, the agents quickly consume all the apples, nothing can regrow, and the tragedy of the commons materialises.

\paragraph{Clean Up}
The reward increases steadily as the number of agents increases, because the apple growth rate is scaled by $\frac{n}{7}$.
We see no collapse, as we saw for Commons Harvest, but this does not mean that the policies are achieving the maximum from the environment: they may not be optimally cleaning the river.

\paragraph{Externality Mushrooms}

\cref{fig:em_pretraining} shows that social welfare increases with the number of players because harvesting mushrooms causes more to spawn; resources are not limited as in Commons Harvest.
However, while social welfare increases with more agents, they are not necessarily achieving the best possible social outcomes: the agents may be underinvesting in the public goods (blue mushrooms).
In the next section, we examine whether reward exchange can improve the social welfare.

\subsection{Training}
\label{sec:training}

\begin{figure}[t]
  \centering
  \begin{subfigure}[t]{\columnwidth}
    \centering
    \includegraphics[width=\columnwidth]{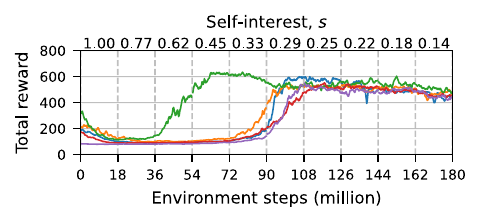}
    \vspace{-4ex}
    \caption{Commons Harvest}
    \label{fig:ch_train}
  \end{subfigure}
  \begin{subfigure}[t]{\columnwidth}
    \centering
    \includegraphics[width=\columnwidth]{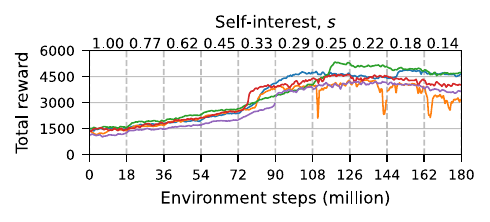}
    \vspace{-4ex}
    \caption{Clean Up}
    \label{fig:cu_train}
  \end{subfigure}
  \begin{subfigure}[t]{\columnwidth}
    \centering
    \includegraphics[width=\columnwidth]{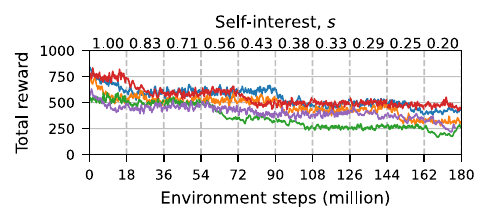}
    \vspace{-4ex}
    \caption{Externality Mushrooms}
    \label{fig:em_train}
  \end{subfigure}
  \vspace{-1ex}
  \caption{Iteratively decreasing self-interest during training}
  \vspace{-2ex}
\end{figure}

Following our method from \cref{sec:method}, we apply reward exchange to the pre-trained policies from \cref{sec:pre_training}, which serve as our challenging policy initialisations.
We iteratively decrease the self-interest of the agents to identify the critical threshold where these initially selfish policies converge to cooperative equilibria.

\paragraph{Commons Harvest}

\cref{fig:ch_train} demonstrates that all seeds have recovered to their maximum performance by $s=0.29$.
Although the agents only achieve a collective reward that is less than what a single agent can achieve, this is due to the difficulty of coordination in independent MARL.
In fact, by $s = 0.14$, the agents have an effective team reward, and they would all be better off if one of them followed the single agent policy learnt in \cref{sec:pre_training} and the other six remained inactive in an empty corner.

We calculate the mean and standard deviation of the total reward achieved at the end of training for all values of $s$, and compute the one-sided Dunnett's test p-value to determine whether the means are statistically lower.
The results are presented in \cref{table:results_ch} where, for brevity, we show only the values of $s$ close to $s^*$.
In this case, the optimally performing value is $s=0.29$.
There is no greater value of $s$ for which the mean social welfare is not statistically worse.
We therefore estimate the self-interest level to be in the range $0.29 \leq s^* < 0.33$.
Note that longer training periods or a larger number of seeds might yield slightly different results. 
However, limited computational resources prevent an exhaustive investigation.

\begin{table}[t]
  \centering
  \begin{subtable}{\columnwidth}
    \centering
    \begin{tabular}{c|ccccc}
      s & 0.46 & 0.33 & \textbf{0.29} & 0.25 & 0.22\\
      \hline
      mean & 208 & 295 & \textbf{540} & 510 & 524\\
      std dev & 231 & 171 & 45 & 48 & 30 \\
      p-value & 0.01 & 0.01 & N/A & 0.17 & 0.27 \\
    \end{tabular}
    \caption{Commons Harvest}
    \vspace{1ex}
    \label{table:results_ch}
  \end{subtable}
  \begin{subtable}{\columnwidth}
    \centering
    \begin{tabular}{c|ccccc}
      s & 0.33 & 0.29 & \textbf{0.25} & 0.22 & 0.18\\
      \hline
      mean & 3646 & 4173 & \textbf{4613} & 4289 & 4313\\
      std dev & 416 & 374 & 439 & 674 & 504\\
      p-value & 0 & 0.06 & N/A & 0.20 & 0.17\\
    \end{tabular}
    \caption{Clean Up}
    \label{table:results_cu}
  \end{subtable}
  \vspace{-1ex}
  \caption{Dunnett's test results}
  \vspace{-2ex}
\end{table}

\paragraph{Clean Up}
Here, we also see that decreasing self-interest leads to an increase in social welfare (\cref{fig:cu_train}), up to a certain point.
When self-interest becomes very low, one of the training seeds experiences instability issues.
\cref{table:results_cu} displays Dunnett's test results.
The value of $s$ with the largest mean again turns out to be the self-interest level.
We therefore estimate the self-interest level to be in the range $0.25 \leq s^* <0.29$.

\paragraph{Externality Mushrooms}
In this environment, reward exchange does not increase social welfare, as shown in \cref{fig:em_train}; performance slightly degrades as self-interest decreases.
These results suggest that Externality Mushrooms does not meet our definition of a Markov social dilemma, as outlined in \cref{sec:markov_sd}, because providing agents with incentives to care about ought to improve collective outcomes.
In \cref{sec:assessment}, we probe the underlying dynamics of this environment and provide further evidence that it is not a Markov social dilemma.

\paragraph{Recordings}
For each environment, we provide a recording of one of the seeds at the end of pretraining and training\footnote{\url{https://github.com/willis-richard/meltingpot/tree/markov_sd/videos}}.
For example, at the end of pretraining in Clean Up, three of the players clean the river at the beginning of the episode.
When apples start spawning, they abandon their chore to harvest the apples, only returning to clean the river after too much pollution accumulates and no more apples appear.
After training with reward exchange, one of these players has learnt to become a full-time cleaner.
This agent harvests no apples, but still gains a proportion of the reward for each apple its co-players harvest.

\subsection{Validation}
\label{sec:validation}
To validate the self-interest level for Commons Harvest and Clean Up, we train new policies continuing from the $n=1, s=1$ policies in \cref{sec:pre_training}, without using a curriculum.
These policies understand the environment dynamics but have not encountered other agents before.
We compare policies trained with $s=1$ (fully independent), $s=s^*$ (self-interest level), $s= \frac{1}{n}$ (team reward) and a value of $s$ slightly larger than the range within which $s^*$ was determined, which we call $s^+$.

\begin{figure}[t]
  \centering
  \begin{subfigure}[t]{\columnwidth}
    \includegraphics[width=\columnwidth]{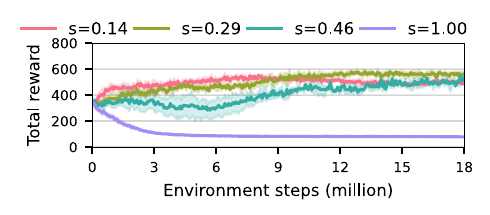}
    \vspace{-4ex}
    \caption{Commons Harvest}
    \label{fig:ch_validate}
  \end{subfigure}
  \begin{subfigure}[t]{\columnwidth}
    \includegraphics[width=\columnwidth]{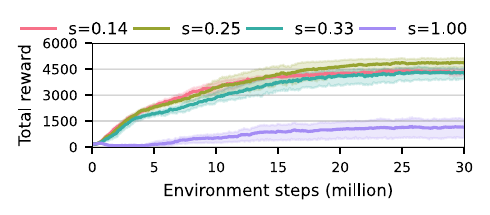}
    \vspace{-4ex}
    \caption{Clean Up}
    \label{fig:cu_validate}
  \end{subfigure}
  \vspace{-1ex}
  \caption{Training without curriculum learning}
  \label{fig:validate}
  \vspace{-2ex}
\end{figure}

Our results in \cref{fig:validate} confirm that the policies do not converge to cooperative equilibria without reward exchange.
The lines show the mean reward received across the seeds, and the shaded regions represent a 95\% confidence interval.
Training at the self-interest level or with a team reward reaches a cooperative equilibrium, as expected.
For both environments, the best performance is achieved with $s^*$.
We posit that training at the game's self-interest level offers an advantage: while socially optimal policies are learnt, the agents retain a greater amount of self-interest than when using a team reward.
This larger individual incentive may alleviate credit assignment difficulties and potential lazy agent issues associated with team rewards.

We expect it to be easier to converge to cooperative equilibria when starting training from single-agent policies, because we do not begin in any particular equilibrium, compared to starting from a selfish equilibrium.
As demonstrated by the $s^+$ runs, which also reach cooperative equilibria, albeit more slowly, it is possible to train cooperative agents with a self-interest level greater than $s^*$.
However, the possibility of converging to a selfish equilibrium may remain.
Therefore, we recommend using the self-interest level to ensure policies can escape such outcomes, as demonstrated in \cref{sec:training}.

\subsection{Assessment}
\label{sec:assessment}

\begin{figure}[t]
  \centering
  \begin{subfigure}[t]{\columnwidth}
    \centering
    \includegraphics[width=\columnwidth]{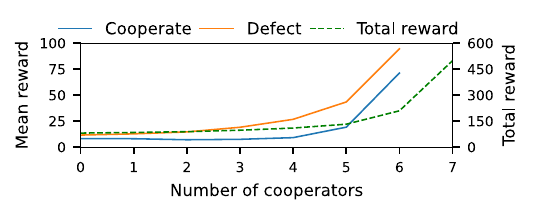}
    \vspace{-4ex}
    \caption{Commons Harvest}
    \label{fig:ch_assess}
  \end{subfigure}
  \begin{subfigure}[t]{\columnwidth}
    \centering
    \includegraphics[width=\columnwidth]{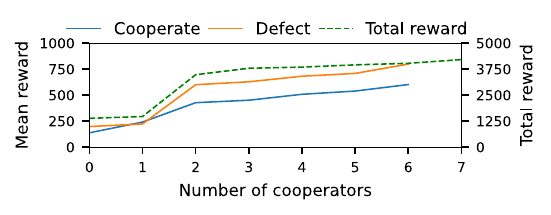}
    \vspace{-4ex}
    \caption{Clean Up}
    \label{fig:cu_assess}
  \end{subfigure}
  \begin{subfigure}[t]{\columnwidth}
    \centering
    \includegraphics[width=\columnwidth]{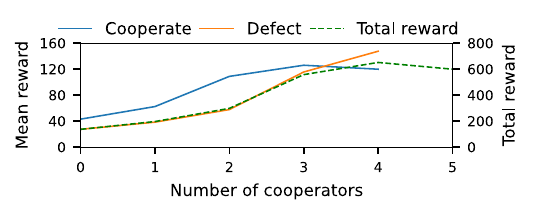}
    \vspace{-4ex}
    \caption{Externality Mushrooms}
    \label{fig:em_assess}
  \end{subfigure}
  \vspace{-1ex}
  \caption{Schelling diagrams}
  \vspace{-2ex}
\end{figure}

To assess whether our environments are Markov social dilemmas, we examine whether they satisfy the inequalities outlined in \cref{sec:markov_sd}.
We select sets of cooperative and defection policies and evaluate the performance of different policy combinations.
Due to the symmetry of our environments, as the agents have homogeneous capabilities and randomised starting locations, the salient point is the average reward received by defecting and cooperating agents, given the number of agents choosing cooperative policies.

We take $n_c$ cooperators from $\Pi_c$ and $n - n_c$ defectors from $\Pi_d$ and evaluate the mean rewards that defect policies and cooperate policies achieve over 225 episodes, for $n_c = 0, 1, \ldots, n$.
We plot these results as a Schelling diagram \cite{schelling73__hockey_helmets_concealed_weapons_and_daylight_saving}, which shows the mean reward to an additional agent depending on whether it chooses to cooperate or to defect, given the number of co-players cooperating.
On the right axis, we also plot the social welfare as a function of the number of cooperators.

\paragraph{Commons Harvest}
We select one of our validation (\cref{sec:validation}) seeds with $s=1$ to represent the defect policies, and one with $s=s^*$ to represent the cooperate policies.
The former typically harvest the last apple in a patch, while the latter are more restrained.
The Schelling diagram is displayed in \cref{fig:ch_assess}.

We conclude that Commons Harvest is a Markov social dilemma because: (1) the social welfare strictly increases as the number of cooperative policies increases; (2) an agent always benefits from choosing a defecting policy compared to a cooperating policy, regardless of how many co-players have chosen to defect; and (3) all agents prefer mutual cooperation to mutual defection, which follows from the game symmetry.

\paragraph{Clean Up}

Again, we select one of our validation seeds with $s=1$ to represent defect policies, and one with $s=s^*$ to represent cooperate policies.
Both the defect policies and cooperate policies tend to clean the river, but the cooperate policies typically start cleaning at a lower pollution level.

Clean Up \cref{fig:cu_assess} is also a Markov social dilemma.
It takes at least two agents simultaneously cleaning the river to reduce the pollution level, which is why we see a jump in social welfare as we move from 1 to 2 cooperators.

\paragraph{Externality Mushrooms}
We select one of our pretraining seeds (\cref{sec:pre_training}) to represent cooperative policies, as these achieve the highest social welfare.
For defect policies, we train new policies with $s=1$ in the environment, without curriculum learning.
The former policies have a greater tendency to harvest prosocial mushrooms compared to the latter.

The Schelling diagram in \cref{fig:em_assess} reveals that this is not a Markov social dilemma because social welfare is maximised when one player defects, violating \cref{eqn:sd_ie_1}.
This helps explain why reward exchange does not improve performance in \cref{sec:training}: exchanging rewards with co-players likely makes it more challenging for an agent to learn the optimal selfish strategy.
In general, the agents are better off choosing cooperative policies.
This result was not immediately apparent from the environment design, as it appeared to incentivise agents to prioritise individual rewards over collective benefits.

Our analysis reveals that harvesting green (prosocial) mushrooms is strategically optimal even for purely self-interested agents.
Although green mushrooms initially provide lower individual rewards than red (selfish) mushrooms, because harvesting a particular mushroom causes more of the same variety to grow, agents ultimately benefit from their co-players' subsequent increased consumption of prosocial mushrooms.

PPO fails to discover this strategic advantage without curriculum learning because the temporal credit assignment challenge is significant: rewards from co-players' subsequent prosocial mushroom consumption occur many timesteps later.
Consequently, the policies almost exclusively harvest selfish mushrooms, converging to an equilibrium with low individual rewards, although agents would individually benefit from switching to cooperative policies.
The discrepancy between our cooperative and defect policies arises because prosocial mushrooms reward agents with $\frac{2}{n}$, which is larger when $n$ is smaller.
Only with curriculum learning, progressively adding more players, does PPO learn to appreciate the long-term benefits of harvesting green mushrooms.



\section{Conclusion}
\label{sec:conclusion}

We introduced a novel method for estimating the self-interest level of Markov social dilemmas, bridging the gap between game-theoretic metrics and complex multi-agent reinforcement learning models.
The self-interest level is a valuable metric for assessing the propensity to cooperate in mixed-motive games by quantifying the gap between individual and collective incentives.

Applying our method to environments from the Melting Pot suite, we determined which environments represent genuine Markov social dilemmas, estimating their self-interest level, and which do not. 
We showed that policies trained with reward exchange at the self-interest level converge to equilibria with good social welfare in Markov social dilemmas (\cref{sec:validation}).

Our work offers both practical metrics and solutions for real-world social dilemmas.
As a metric, the self-interest level enables risk assessment: systems with low self-interest levels face significant barriers to cooperation and may be prone to conflict, allowing system designers to identify where intervention is necessary.
Additionally, assessing the self-interest level can provide insights into why certain reinforcement learning algorithms perform better or struggle in specific environments, aiding practitioners in selecting appropriate algorithms for mixed-motive scenarios.
As a solution mechanism, reward exchange can be applied to problems like fishery management, where traditional quotas often fail due to persistent incentives to overfish.
If fishing nations were to exchange a proportion of their fishing profits with other fishing nations, this would reduce each country's incentive to overexploit fish stocks while simultaneously motivating all participating countries to improve ocean health.

Future work in this area could include developing a method to determine the general self-interest level \cite{willis24__resolving_social_dilemmas_with_minimal_reward_transfer} for Markov social dilemmas.
This generalisation allows agents greater freedom in their reward transfers, which can improve efficiency in asymmetrical environments.
For example, if there are multiple types of agents, we could use separate parameters to govern the proportion of reward exchanged between agents of different types.
Furthermore, the experimental approach in this paper could be applied to more environments, such as those in the Melting Pot suite, to delve into their underlying dynamics and challenges.
We could investigate the impact on the self-interest level of systematic changes to an environment, for example, by increasing resource scarcity or introducing a sanctioning mechanism.

\section*{Acknowledgments}
This work was supported by UK Research and Innovation [grant number EP/S023356/1], in the UKRI Centre for Doctoral Training in Safe and Trusted Artificial Intelligence (\url{www.safeandtrustedai.org}) and a BT/EPSRC funded iCASE Studentship [grant number EP/T517380/1].

Compute resources were provided by King's College London \cite{kingscollegelondone-researchteam24__kings_computational_research_engineering_and_technology_environment_create}.

\bibliographystyle{named}
\bibliography{library}

\end{document}